\documentclass[aip,apl,graphicx, reprint]{revtex4-1}
\usepackage{graphicx,amsmath,psfrag,mathrsfs}
\draft 

\begin{document}


\title{Operation of transition edge sensors in a resistance locked loop} 



\author{J. van der Kuur}
\email{j.van.der.kuur@sron.nl}
\affiliation{SRON national institute for space research, Sorbonnelaan 2, 3584CA Utrecht, the Netherlands}
\author{M. Kiviranta}
\affiliation{VTT, Tietotie 3, 02150 Espoo, Finland }


\date{23 november 2012}

\begin{abstract}
We propose to operate a superconducting transition edge sensor (TES) using a different type of biasing, in which the resistance of the TES is kept constant by means of feedback on the bias voltage and is independent of the incoming signal power. By combining a large negative electrothermal feedback with a load independent resistance, this approach can significantly linearise the response of the detector in the large signal limit. The electrothermal feedback is enhanced in comparison with the commonly applied voltage biasing, which further increases the speed of the detector. Furthermore, in frequency domain multiplexed (FDM) readout, the sinusoidal bias voltages for each TES can be generated cryogenically with the readout SQUIDs. 
\end{abstract}


\maketitle 

Superconducting transition edge sensors (TES) are widely used for cryogenic thermal radiation detection in a large variety of wave lengths, ranging from submm \cite{jackson12} to gamma ray detectors \cite{irwin95,ullom08}.  TES detectors exploit the steep resistive phase transition of superconductors at their transition temperature $T_c$ to translate applied optical signals into resistance changes.  Both a linear response function and a fixed response time are very important for convenient operation of TES-based instruments. 
In order to achieve linear behaviour in the intrinsically nonlinear TES-based detectors, negative electrothermal feedback\cite{irwin95,irwinhilton05} (ETF) with a high loop gain has been successfully applied for many years to stabilise the operating temperature at a fixed value. This negative feedback is created by a combination of stiff voltage bias and the steep resistive transition of a TES.  This way of operation creates a stable dynamic equilibrium between the Joule heating power $P_{J_0}$ dissipated in the TES on one hand,  and cooling to the thermal bath at temperature $T_b$ minus the applied signal power on the other hand, at a steady state operating temperature of $T_0>T_b$ with $T_0\approx T_c$. When $T_b$ is kept constant,  the feedback signal provides the required Joule heating power to keep the detector in its operating temperature. Applied optical signals to the detector lower the required heating power, which in turn is mirrored in a lower feedback signal power which forms the detected signal.

In this paper we will summarise the generally unattractive detector properties for large applied signals, which are obtained when stiff voltage is used for implementation of strong negative ETF. We will present an alternative way of obtaining strong negative ETF, which provides virtually signal level independent detector properties instead. By stabilising the TES resistance using feedback, a combination of strong negative ETF and a constant TES resistance is obtained.  This approach is specifically attractive in the case of sinusoidal (AC) TES bias and frequency domain multiplexing (FDM), as it provides a way to use the readout SQUIDs as TES bias generators.  

For small thermal signals referred to $P_{J_0}$, negative ETF as a result of voltage bias linearises the response to signal power and makes it only dependent on the bias voltage, and independent of the steepness of the transition for frequencies well below the thermal roll-off  of the detector. Furthermore, in the limit of stiff voltage bias, the ETF reduces the thermal time constant of the detector from $\tau_0=C/G$ to 
\begin{equation}
\tau_{\mathrm{eff}} = \tau_0 \frac{1+\beta_I}{1+\beta_I +\mathscr{L}_I},
\end{equation}
where $\mathscr{L}_I \equiv P_{J_0} \alpha_I/G T_0$ with $P_{J_0} = I_0^2R_0$ defines the low-frequency loop gain under constant current of the electrothermal feedback,  $C$ the heat capacity of the detector, $G$ the local derivative of the thermal conductivity between the detector and the heat bath, and $I_0$, $R_0$ the steady state bias current and resistance of the TES, respectively. Furthermore, two dimensionless parameters $\alpha_I \equiv \partial \log R/\partial \log T |_{I_0}$ and $\beta_I \equiv \partial \log R/\partial \log I |_{T_0}$ define the steepness of the TES transition with $T$, $I$, and $R(T,I)$ the instantaneous temperature, bias current, and resistance of the TES, respectively. Note that the thermal time constant of the detector is directly proportional to the $\alpha_I(R)$ and $\beta_I(R)$ parameters, and therefore depends significantly on the local properties of the transition. 

In the large signal limit, i.e. when the signal power becomes a significant fraction of the Joule heating power $P_{J_0}$, the loaded detector under voltage bias equilibrates at a higher temperature and hence resistance, with usually lower local $\alpha_I(R)$ and $\beta_I(R)$ parameters. 
This large signal behaviour under ETF with stiff voltage bias possesses a number of properties which are generally unattractive for applications. First, the detector slows down significantly under large signal power loadings. In applications where the speed of the detector is an integral part of the calibration accuracy, such as in the combination of an FTS with TES-based bolometers in the Safari instrument\cite{jackson12} for the SPICA mission, this effect limits the overall performance of the instrument. Secondly, the saturation power of the detector  depends on the set point under unloaded conditions, so that a compromise has to be made between small signal properties and saturation power. Thirdly, the changing TES resistance under optical loading reduces the efficiency with which the bandwidth can be used in multiplexing schemes\cite{irwin02}, as the consumed bandwidth scales with the TES resistance. 

\begin{figure}
\includegraphics[width=0.8\columnwidth]{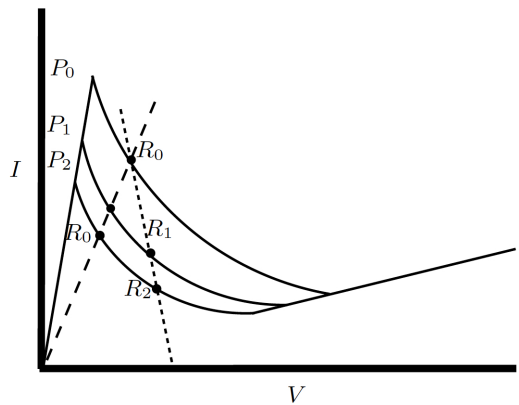}
\caption{Sketch of the $IV$ curves of a TES-based bolometer under different optical loadings ($P_2 < P_1 < P_0$), as well as the load lines of the different bias sources. The dotted line shows the bias points under different optical loadings for practical non-ideal voltage bias, while the dashed line shows the bias points for biasing in the resistance locked loop (RLL). Under voltage bias the equilibrium resistance $R_{0..2}$ changes under optical loading, whereas in RLL bias the TES resistance remains constant.}
\label{fig:schema}
\end{figure}

These inconvenient large signal properties of TES detectors can be avoided if the resistance of the detector is kept constant under signal power loading, while maintaining a large negative ETF.  This mode of operation, which we will refer to as the resistance locked loop (RLL), is illustrated in Fig.~\ref{fig:schema} where the $IV$-curves of TES-based bolometer are sketched under different optical loadings. Each curve is almost hyperbolic, as the width of the superconducting TES transition is generally very small with respect to the critical temperature $T_c$. As a result, the Joule heating $P_{J_0}$ which is required to keep the TES in the transition, is virtually independent of the operating point in the transition, so that $P_{J_0}=I_0V_0=\mathrm{constant}$ with $V_0$ being the steady state voltage drop across the TES. Under optical loading the decrease in Joule heating matches the optical power.

The $IV$-curve labeled $P_0$ refers to the situation without optical loading, and the labels $P_1$ and $P_2$ refer to the curves under increasingly higher optical loadings. The dotted line shows the load line of a practical, non-stiff voltage bias source. The crossings between the $IV$-curves and the load line (labeled $R_0$, $R_1$, and $R_2$), indicate the resulting observed resistances under optical loading. As the resistances change significantly under optical power loading, the TES parameters $\alpha_I(R)$ and $\beta_I(R)$ change significantly, and therefore the thermal time constant of the detector.

The RLL implies that a trajectory should be followed as indicated by the dashed line in Fig.~\ref{fig:schema}, which intersects with the origin. We propose to operate the detector following this load line, by feeding back the measured current to the applied bias voltage in such a way that the ratio between the applied voltage $V_0$ and observed steady state current $I_0$, i.e. the resistance $R_0=V_0/I_0$, remains constant independent of signal power loading (i.e. a RLL).  As the resistance is kept independent of loading, so are $\alpha_I(R)$ and $\beta_I(R)$  which govern the detector roll-off, and the consumed bandwidth $R_0/2\pi L$, where $L$ is the inductance in the bias circuit. Furthermore, the saturation power of the detector becomes independent of the bias point.

Feedback of the measured current to the bias voltage has been demonstrated earlier\cite{nam99} for the purpose of reducing the thermal response time of micro-calorimeters to enhance their photon count rate capability. The case of a load line through the origin, however, was considered to be unstable as a result of an infinite ETF loop gain. However, the $\beta_I(R)$ was not included in the model. We have found that when a non-zero $\beta_I(R)$ parameter is taken into account, the scheme is stable and provides a large but finite negative ETF loop gain, in combination with the set of attractive properties associated with a constant resistance.

As the scheme requires feedback on the TES bias voltage it is easily applicable in FDM, as all bias voltages are separated in frequency space and multiplexed like the detector signals\cite{kuur03}. In the case of time domain multiplexing (TDM), however, the RLL scheme would require a bias wire per pixel, as the TES bias voltages cannot be multiplexed since they overlap in frequency and phase space. 

In order to understand the the behaviour of the detector under the RLL bias conditions, we use the detector small signal models (assuming DC bias) which have been elaborated elsewhere\cite{irwinhilton05}.  Note that under sinusoidal (AC) bias the same model is applicable\cite{kuur11}, when substituting the root mean square (RMS) values of the voltages and currents for $V_0$ and $I_0$, respectively, and $2L$ for the bias circuit inductance.

In the RLL, the applied voltage to the TES by the bias source in the small signal limit in the time domain can be described as
\begin{equation*}
\left.
\begin{array}{l}
V(t) = I(t)  \cdot R_0 \\ I(t) = I_0+\Delta I(t)
\end{array}
\right\}
\Rightarrow V(t) = \underbrace{I_0 R_0}_{V_0} - (-R_0) \Delta I(t) .
\end{equation*}
The resulting small signal Th\'{e}venin equivalent of this bias source consists of an ideal voltage source $V_0=I_0 R_0$, with a negative resistance $-R_0$ in series.

Such a bias source cannot be constructed from passive components, but can be created actively using feedback. As a result, inevitable propagation delay which is present in the feedback circuit imposes a bandwidth limitation on the feedback signal which explicitly needs to be taken into account to obey stability conditions. Taking these factors into account, we obtain
\begin{equation}
V(s) = V_0 \delta(s=0) -  \frac{-R_0}{1 + s \tau_d } \cdot I(s)
\label{eq:rllsrc}
\end{equation}
as transfer function for the bias source in the frequency domain, with $-R_0$ the internal negative resistance of the bias source, $s=j\omega$, and $s \tau_d$ representing the dominant pole in the RLL feedback circuit. 

We now can include the behaviour of the RLL by substitution of eq.~(\ref{eq:rllsrc}) in the standard small signal model in the frequency domain, obtaining
\begin{multline*}
\begin{pmatrix}  v_L +v_J \\ P_{opt}(s) +P_{ph} -I_0 v_J \end{pmatrix} 
=  \\
\begin{pmatrix} Z_L+R_0(1+\beta_I)\underbrace{-\frac{R_0}{1+s\tau_d}}_{\mathrm{RLL}} & 
	G \mathscr{L}_I/I_0  \\ 
	-V_0(\beta_I+2) & G(1-\mathscr{L}_I +s\tau_0) \end{pmatrix} 
 \begin{pmatrix} \Delta I(s)  \\ 
         \Delta T(s) \end{pmatrix}
\end{multline*} 
In these equations $Z_L=R_L + sL$ is the impedance of the TES bias circuit consisting of a load resistor $R_L$ and an inductor $L$. Furthermore, $P_{opt}(s)$ is the  absorbed optical signal power, $P_{ph}=\sqrt{4\gamma kT^2G}$ the phonon noise spectral density of the thermal link $G$ with $0 < \gamma \leq 1$ a geometry dependent factor\cite{mather82}, and $v_L$  and $v_J$ are the Johnson noise spectral densities of the load resistor $R_L$, and the TES steady state resistance $R_0$, respectively.

This set of equations can be solved rigorously for $\Delta I(s)$ both analytically and numerically. However, doing so will not give more insight than the derivation shown below, which involves a few simplifying, but realistic, assumptions.

As under most operating conditions the gain-bandwidth-product ($1/2\pi\tau_d $) of the RLL can be much higher than the electrical bandwidth ($\approx R_0(1+\beta_I)/2\pi L$) of the detector, we can approximate the RLL transfer function by
\begin{equation}
\frac{R_0}{1+s \tau_d} \approx R_0(1-s \tau_d).
\label{eq:taux}
\end{equation}
To ensure electrothermal stability the electrical circuit is often dimensioned to be much faster than the thermal circuit (i.e. $\tau_{\mathrm{eff}} \gg L/(R_0(1+\beta_I)) \gg \tau_d$). We therefore omit the pole resulting from the electrical circuit to simplify the interpretation of the expressions.
Under these assumptions, expressions are obtained for the current spectral densities which are very similar to the standard voltage bias case. 
For the responsivity $\Delta I_{th}(s)/\Delta P(s)$, with $\Delta P(s)$ the frequency dependent power input to the detector, we find 
\begin{equation*}
\frac{\Delta I_{th} (s)}{\Delta P(s)} = - \frac{1}{2 V_0} \cdot \frac{\mathscr{L}_I ^{'}}{\mathscr{L}_I ^{'} +1} \cdot \frac{1}{1+s\tau_\mathrm{eff}^{'}},
\end{equation*}
with
\begin{equation*}
\mathscr{L}_I ^{'} = \frac{2 \mathscr{L}_I }{\beta_I} = \frac{2}{\beta_I} \frac{P_0 \alpha_I}{G T_0} \propto \frac{\alpha_I}{\beta_I},
\end{equation*}
and $\tau_{\mathrm{eff}}^{'}=\tau_0/(1+\mathscr{L}_I ^{'})$. 
The TES Johnson noise spectral density contributes
\begin{equation*}
\Delta I_J(s) = \frac{v_J}{\beta_I R_0}\cdot  \frac{1}{1+\mathscr{L}_I ^{'}} \cdot \frac{1+s\tau_0}{1+s\tau_e^{'}},
\end{equation*}
to the observed current spectral density, and the load resistor Johnson noise produces
\begin{equation*}
\Delta I_L(s) = \frac{v_L}{\beta_I R_0} \frac{1-\mathscr{L}_I }{1+\mathscr{L}_I ^{'}} \cdot \frac{1+s\tau_I}{1+s\tau_e^{'}},
\end{equation*}
with $\tau_I \equiv \tau_0/(1-\mathscr{L}_I )$.

Note that the expressions for the responsivity,  the TES Johnson noise, the load resistor noise, and the impedance can be obtained by substitution of $\beta_I \rightarrow (\beta_I -2)/2$ and $V_0 \rightarrow 2 V_0$ in the well known voltage biased expressions.

A number of different properties are obtained with respect to the stiff voltage bias operation. First, the effective fall time of the detector decreases by a factor $2+2/\beta_I$ in the limit of $\mathscr{L}_I \gg 1$, which is attractive for bolometers\cite{jackson12}. Secondly, the responsivity in the current domain diminishes with a factor of 2, as the other half the signal power screening is observable in the voltage domain. Note that as in any feedback scheme, the signal-to-noise ratio does not change in the RLL. As a result of that, the fundamental NEP in bolometers or energy resolution of TES-based micro-calorimeters does not change in the RLL operation. 


\begin{figure}
\centering
\includegraphics[width=0.8\columnwidth]{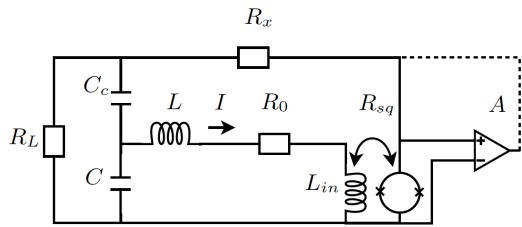}
\caption{Schematic diagram of an implementation of the RLL. A voltage $V(t) = I(t) R_0$ is created proportional to the measured current $I(t)$, when $R_{sq} A R_L C_c/R_x C= R_0$ (in the limit of large voltage division ratios and $R_L \ll R_0$). The solid lines show an implementation at cryogenic temperature, and the dotted line shows a feedback loop through room temperature electronics.}
\label{fig:imp}
\end{figure}

DC and AC voltage biased cryogenic TES detectors are commonly readout with SQUID trans impedance amplifiers with the linear fraction of their trans resistance equal to $R_{sq} \equiv V_{sq}/I$. Their voltage bias sources generally consist of a voltage source at room temperature in series with a voltage divider with a large division ratio for noise reasons. 

Fig.~\ref{fig:imp} shows two possible implementations of the RLL in an AC biased TES detector ($R_0$).   The bias voltage divider consists of a resistive divider $R_L/(R_x+R_L)$ in combination with a capacitive voltage divider $C_c/(C+C_c)$. The capacitors $C$ and $C_c$ in combination with inductors $L$ and $L_{in}$ tune the bias frequency of the circuit to $\omega_0=1/\sqrt{(L+L_{in})(C+C_c)}$.

The RLL is created by feeding back the output voltage of the  SQUID amplifier $V_{sq}$ through the bias voltage divider, in such a way that the ratio between the net applied TES bias voltage and the TES bias current $I$ becomes equal to the required TES bias resistance value $R_0$, i.e. when $R_{sq} A R_L C_c/R_x C= R_0$ (in the limit of large voltage division ratios and $R_L \ll R_0$). The output voltage of the SQUID amplifier $V_{sq}$ is either fed back directly at cryogenic temperatures (solid line, $A \equiv 1$), or at room temperature (dotted line, $A>1$). 

Note that in case of DC biased TES detectors, the offset voltage of the SQUID resulting from its bias current cannot be blocked by a capacitor, but needs to be cancelled actively by a compensation network.

The nonlinear fraction of the transfer function of the SQUID will create harmonics, of which the amplitude depends on the net  flux excursion at the SQUID. Linearisation techniques\cite{kiviranta08} can be used to minimise these effects to the required levels for the application.

We identify another interesting property of the AC version of the RLL biasing scheme.  The loop gain in the RLL equals
\begin{equation*}
\frac{R_0}{1+s\tau_d} \frac{1}{sL+1/sC+Z_{tes}} = -1 
\end{equation*}
when $\omega=1/ \sqrt{LC}$, the TES impedance\cite{irwinhilton05} $Z_{tes} \approx -R_0$, and $\omega\tau_d \ll 1$, i.e. when the denominator is real. This implies that the circuit obeys the Barkhausen criterion for oscillators, which implies that the oscillation is self sustaining. The circuit behaves as an oscillator modulated by absorbed optical power, of which the frequency is set by the $LC$ filter resonance frequency, and the amplitude by the bias current which is required to keep the TES at its bias resistance $R_0$ under the applied signal loading conditions. The electrothermal feedback in the TES stabilises the amplitude in the same way as the incandescent lamp does in a filament stabilised Wien bridge oscillator\cite{meacham38}. 

As a result, no external bias source is necessary to keep the TES detector in its operating point, as the bias power is provided by the SQUID amplifier.  When multiple resonating TES bias circuits are connected to the same SQUID as in frequency domain multiplexed (FDM) TES readout\cite{kuur03}, the SQUID will provide bias power to all TES bias circuits independently.  The multiplexing factors which can be achieved with this scheme are expected to be considerable. For example, in case of an array of bolometers with $P_{J_0}=10$~fW and $T_c=100$~mK,  we expect that a multiplexing factor of $N\approx 100$ can be achieved with a SQUID with a flux noise of $0.25 \mu\phi_0/\sqrt{\mathrm{Hz}}$, based on dynamic range considerations.  The maximum stable resonance frequency of a self-biased (AC) TES circuit is limited by the propagation delay in the feedback circuit to $\omega \tau_d \lesssim 0.1$. 

The dynamic range requirements for the SQUID decrease in the RLL, as the SQUID amplifier is in the forward path of a closed feedback circuit. The flux noise of the SQUID acts upon the circuit in the same way as the load resistor noise $v_L$, and is therefore suppressed by electrothermal feedback.  As a result, the effective flux noise as observed at the SQUID output will be decreased by a factor $\beta_I/(1+\beta_I)$. Consequently, for operating points close to the normal state, where the value of $\beta_I$ commonly is considerably smaller than one, the required SQUID dynamic range relaxes with a factor of $0.5+1/2\beta_I$. Note that this effect has not been included in the estimations for the multiplexing factor mentioned above.

In summary, we have shown that when a TES-based detector is biased in a resistance locked loop, it operates stable and with an enhanced ETF loop gain, speed, and improved linearity for large signals. The fixed operating resistance offers important improvements over the standard operating conditions in the voltage bias, specifically in the large signal limit and in multiplexing. Furthermore, the TES bias power can be provided directly by the SQUID amplifier, instead of from sources at room temperatures. This significantly simplifies the implementation of frequency domain multiplexing. Finally, the SQUID dynamic range requirements decrease for small values of $\beta_I$, as the SQUID is in the forward path of the feedback circuit. Practical demonstration of the operating mode is ongoing.

The authors thank Luciano Gottardi and Pourya Khosropanah, and Jian-Rong Gao for useful discussions.


%
%

%


\bibliography{lit-lst}

\end{document}